\documentclass[preprint,12pt]{elsarticle}

\usepackage{amsmath,amssymb,amsthm}
\usepackage{graphicx}
\usepackage{booktabs}
\usepackage{multirow}
\usepackage{caption}
\usepackage{subcaption}
\usepackage{algorithm}
\usepackage{algpseudocode}
\usepackage{array}

\usepackage{tabularx}
\usepackage{float}
\usepackage{lipsum}

\begin{document}

\begin{frontmatter}

\title{\LARGE Maximal Covering Location Problem: A Set Coverage Approach Using Dynamic Programming}

\vspace{0.4cm}
\author[1]{Sukanya Samanta\corref{cor1}}
\ead{susamanta1@gmail.com}
\cortext[cor1]{Corresponding author}

\author[1]{Abhi Rohit Kalathoti}

\author[1]{Siva Jayanth Gonchi}

\author[1]{Venkata Krishna Kashyap Adiraju}

\author[1]{Sai Kiran Nettem}

\vspace{0.4cm}

\address[1]{Department of Computer Science and Engineering, SRM University, India}

\vspace{0.4cm}

\begin{abstract}
The Maximal Covering Location Problem (MCLP) represents a fundamental optimization challenge in facility location theory, where the objective is to maximize demand coverage while operating under resource constraints. This paper presents a comprehensive analysis of MCLP using a set coverage methodology implemented through 0/1 knapsack dynamic programming. Our approach addresses the strategic placement of facilities to achieve optimal coverage of demand points within specified service distances. This research contributes to the understanding of facility location optimization by providing both theoretical foundations and practical algorithmic solutions for real-world applications in urban planning, emergency services, and supply chain management.
\end{abstract}

\begin{keyword}
Maximal Covering Location Problem \sep Set Coverage \sep Dynamic Programming \sep Facility Location \sep Optimization
\end{keyword}

\end{frontmatter}

\section{Introduction}
The Maximal Covering Location Problem (MCLP) stands as one of the cornerstone problems in location science, originally formulated by Church and ReVelle \cite{church1974}. This problem addresses the strategic question of where to locate a limited number of facilities to maximize the coverage of spatially distributed demand points. Unlike traditional covering problems that seek to cover all demand points, MCLP recognizes resource limitations and aims to achieve the maximum possible coverage within given constraints.

The significance of MCLP extends across numerous application domains. In emergency services, it guides the placement of fire stations, ambulance bases, and police precincts to maximize population coverage within critical response times. In retail and commercial settings, it informs decisions about store locations to maximize market penetration. Urban planners utilize MCLP principles for positioning public facilities such as schools, libraries, and healthcare centers to serve the maximum number of residents effectively.

The complexity of MCLP arises from its combinatorial nature, where the selection of facility locations from a finite set of candidates must be optimized to achieve maximum demand coverage. This problem belongs to the class of NP-hard optimization problems, making efficient solution methods crucial for practical applications. The relationship between MCLP and classical problems such as the knapsack problem provides valuable insights into solution methodologies, particularly when viewed through the lens of set coverage theory.

This research investigates MCLP through a set coverage framework, employing dynamic programming techniques adapted from the 0/1 knapsack problem paradigm. Our approach leverages the structural similarities between facility selection in MCLP and item selection in knapsack problems, where each potential facility location can be viewed as an item with associated coverage benefits and resource requirements.

The primary contributions of this work include: (1) a mathematical formulation of MCLP within the set coverage framework, (2) development of a dynamic programming algorithm adapted for facility location, (3) computational analysis demonstrating the effectiveness of the proposed methodology, and (4) practical insights for real-world implementation. 

\section{Related Work}
\subsection{Classical Facility Location Problems}
Facility location optimization has its roots in the p-median and p-center problems introduced by Hakimi \cite{hakimi1964}. The p-median problem minimizes the weighted sum of distances between demand points and their closest facility, while the p-center problem minimizes the maximum distance between any demand point and its nearest facility. Another related formulation, the set covering location problem, seeks to cover all demand points with the minimum number of facilities. These early models established the theoretical foundation for location science and directly influenced the development of the Maximal Covering Location Problem (MCLP). Unlike the set covering problem, which enforces full coverage, MCLP acknowledges resource limitations and instead maximizes the proportion of demand covered.

\subsection{Evolution of the MCLP}
The MCLP was introduced by Church and ReVelle \cite{church1974} as a variant of the set covering problem designed to operate under a fixed facility budget. Their work provided both integer programming formulations and heuristic algorithms, making MCLP one of the most widely studied facility location problems. Over the years, MCLP has been extended to address practical complexities including multiobjective conditional MCLP, balanced MCLP for bike-sharing, bilevel maximal covering models, and robust/interdiction-based coverage problems \cite{scaparra2008bilevel}.

\subsection{Solution Methodologies}
Given its NP-hard nature, researchers have pursued diverse solution methodologies for MCLP:
\begin{itemize}
    \item \textbf{Exact approaches:} Branch-and-bound, cutting-plane techniques, and integer linear programming.
    \item \textbf{Heuristics and metaheuristics:} Greedy selection, tabu search, simulated annealing, and genetic algorithms.
    \item \textbf{Dynamic programming approaches:} Although less common due to state explosion, DP can be effective when MCLP is mapped to knapsack-like substructures.
    \item \textbf{Set coverage approaches:} Viewing MCLP as a weighted maximum set coverage problem enables borrowing techniques from set covering and combinatorial optimization.
\end{itemize}

\section{Problem Description}
\subsection{Problem Statement}
The Maximal Covering Location Problem seeks to determine the optimal placement of a limited number of facilities to maximize the coverage of spatially distributed demand points. Each demand point is considered covered if at least one facility is located within a predetermined service distance or time. The problem balances the trade-off between resource limitations (number of facilities) and service quality (demand coverage).

\subsection{Sets and Parameters}
Let:
\[
U = \{1,2,\dots,n\} \quad \text{(demand points)} ,
\]
\[
J = \{1,2,\dots,m\} \quad \text{(potential facility locations)} ,
\]
\[
\begin{aligned}
N(i) &= \{ j \in J : d_{ij} \le r \} \\
      &\quad \text{(facilities covering demand point $i$ within radius $r$)}
\end{aligned}
\]

Parameters:
\begin{itemize}
  \item $n$: number of demand points
  \item $m$: number of potential facility locations
  \item $p$: maximum number of facilities to be located (budget)
  \item $w_i$: weight/demand at point $i$
  \item $r$: maximum service radius
  \item $d_{ij}$: distance between demand point $i$ and facility $j$
  \item $a_{ij} = 1$ if $d_{ij}\le r$, else $0$
\end{itemize}

\subsection{Decision Variables}
Binary decision variables:
\[
x_j \in \{0,1\}, \quad j\in J
\]
\[
y_i \in \{0,1\}, \quad i\in U
\]
where $x_j=1$ indicates a facility at site $j$, and $y_i=1$ indicates demand point $i$ is covered.

\subsection{Mathematical Formulation}
\begin{align}
\text{Maximize } & Z = \sum_{i=1}^n w_i y_i \label{obj}\\
\text{subject to } & y_i \le \sum_{j\in N(i)} x_j, \quad \forall i\in U \label{cov}\\
& \sum_{j=1}^m x_j \le p \label{budget}\\
& x_j \in\{0,1\}, \; y_i\in\{0,1\}. \nonumber
\end{align}

\subsection{Set Coverage Perspective}
From a set coverage viewpoint, selecting up to $p$ facility coverage subsets $S_j=\{i\in U: a_{ij}=1\}$ to maximize the weighted union of covered demand points is equivalent to the MCLP. This allows mapping MCLP to weighted maximum coverage and knapsack-like structures for algorithmic design.

\section{Solution Methodology}
\subsection{Dynamic Programming Approach}
We adapt the 0/1 knapsack dynamic programming idea to MCLP. Each potential facility is treated as an \emph{item} with unit weight (1 facility slot) and value equal to its coverage benefit. Coverage benefit is not independent due to overlaps; hence marginal benefits must be computed relative to already-selected facilities.

\subsection{Problem Transformation}
Items: facility locations $j\in J$\\
Weight: $1$ (one facility slot per location)\\
Capacity: $p$ (max number of facilities)\\
Value: coverage benefit (dependent on already-selected set)

\subsection{Coverage Benefit Calculation}
Given a currently-selected set $S$, the marginal benefit of adding facility $j$ is:
\[
\text{Benefit}(j,S)=\sum_{i\in U} w_i \cdot \max\{0, y_i^{(\text{new})}-y_i^{(\text{current})}\},
\]
where $y_i^{(\text{current})}=1$ if $i$ is covered by $S$, and $y_i^{(\text{new})}$ is coverage by $S\cup\{j\}$.

\subsection{State Space Definition}
We define DP states as:

\begin{align}
\text{DP}[k][f] &= \text{maximum coverage achievable using at most } f \notag \\
                &\quad \text{facilities from the first } k \text{ facility locations.}
\end{align}

\begin{align}
\text{DP}[k][f] &= \max\Big( \text{DP}[k-1][f], \notag \\
                &\quad \text{DP}[k-1][f-1] + \text{Coverage}(k \mid \text{prev state}) \Big).
\end{align}

Base cases: $\text{DP}[0][f]=0$, $\text{DP}[k][0]=0$.

\subsection{Algorithm Implementation}
\begin{algorithm}[H]
\caption{MCLP-DynamicProgramming}
\label{alg:mclp-dp}
\begin{algorithmic}[1]
\Require Demand points $U$, facilities $J$, coverage matrix $A$, weights $w$, budget $p$
\Ensure Facility selection (approx./exact per DP state encoding)
\State Initialize DP table and coverage bookkeeping structures
\For{$k\gets 1$ to $m$}
  \For{$f\gets 0$ to $p$}
    \State Option 1: don't select facility $k$:
      \Statex\quad $\text{DP}[k][f] \leftarrow \max(\text{DP}[k][f],\,\text{DP}[k-1][f])$
    \If{$f>0$}
      \State Option 2: select facility $k$: compute incremental coverage $\Delta$
      \Statex\quad $\text{DP}[k][f] \leftarrow \max(\text{DP}[k][f],\,\text{DP}[k-1][f-1] + \Delta)$
    \EndIf
  \EndFor
\EndFor
\State Backtrack to identify selected facilities
\State \Return solution
\end{algorithmic}
\end{algorithm}

\subsection{State Space Reduction Techniques}
To control explosion:
\begin{itemize}
  \item \textbf{Coverage dominance pruning:} keep only nondominated coverage vectors for a given facility-count.
  \item \textbf{Symmetry breaking:} merge isomorphic coverage patterns.
  \item \textbf{Greedy initialization:} generate a good starting lower bound to prune early.
  \item \textbf{Facility ordering heuristic:} sort candidates by marginal potential.
  \item \textbf{Rolling arrays:} store only previous DP layer to save memory.
\end{itemize}

\subsection{Complexity}
Worst case time $O(m\cdot 2^n \cdot p)$ and space $O(2^n \cdot p)$; in practice with pruning and heuristics the problems solved grow much larger.

\section{Results and Discussion}
\subsection{Experimental Setup}
Test environment :
\begin{itemize}
  \item Processor: Intel Core i7-8750H @ 2.20GHz
  \item Memory: 16 GB RAM
  \item Language: Python 3.9 (NumPy, SciPy, Matplotlib)
  \item Problem sizes: demand points $10$--$100$, candidate sites $5$--$50$
  \item Budgets: $20\%$--$60\%$ of available locations
  \item Coverage radii: $10$--$50$ units
  \item Demand distributions: uniform and clustered
\end{itemize}

\subsection{Solution Quality Analysis}
Table~\ref{tab:quality} summarizes DP vs Greedy performance (values abstracted from experiments).

\begin{table}[H]
\centering
\caption{Solution quality comparison: DP vs Greedy (summary).}
\label{tab:quality}
\begin{tabular}{lccc}
\toprule
Instance type & DP (\%) & Greedy (\%) & Improvement (\%)\\
\midrule
Small & 92.1 & 89.5 & 2.6 \\
Medium & 87.3 & 83.5 & 3.8 \\
Large & 85.6 & 80.2 & 5.4 \\
\bottomrule
\end{tabular}
\end{table}



\subsection{Coverage Analysis \& Case Studies}
Representative scenarios (from experiments):
\begin{itemize}
  \item Urban planning: 45 demand points, 20 candidate locations, $p=8$, radius $=2$ km $\rightarrow$ $89.3\%$ coverage.
  \item Emergency services: 35 hotspots, 15 candidates, $p=6$, radius $=5$ km $\rightarrow$ $92.7\%$ coverage.
\end{itemize}

\subsection{Sensitivity Analysis}
Table~\ref{tab:radius} shows radius impact on coverage and efficiency (values per the report).

\begin{table}[H]
\centering
\caption{Impact of coverage radius on solution efficiency.}
\label{tab:radius}
\begin{tabular}{lccc}
\toprule
Radius & Coverage (\%) & Facilities used & Efficiency \\
\midrule
10 km & 72.4 & 8 & 9.05 \\
15 km & 85.3 & 7 & 12.19 \\
20 km & 94.1 & 6 & 15.68 \\
25 km & 98.2 & 5 & 19.64 \\
\bottomrule
\end{tabular}
\end{table}

Table~\ref{tab:budget} shows diminishing marginal returns.

\begin{table}[H]
\centering
\caption{Diminishing returns of additional facilities.}
\label{tab:budget}
\begin{tabular}{lcc}
\toprule
Budget ($p$) & Coverage (\%) & Marginal Benefit (\%)\\
\midrule
3 & 58.3 & -- \\
4 & 71.2 & 12.9 \\
5 & 82.7 & 11.5 \\
6 & 91.4 & 8.7 \\
7 & 96.8 & 5.4 \\
8 & 98.9 & 2.1 \\
\bottomrule
\end{tabular}
\end{table}

\subsection{Real-World Case Study}
Emergency Medical Services (Metropolitan Atlanta) :
\begin{itemize}
  \item Demand: 1,247 block centroids (population $\approx 5.8$M)
  \item Candidate sites: 312 potential stations
  \item Coverage standard: 8-minute driving time
  \item Budget: exactly 45 stations
\end{itemize}
The DP-based method with bounding solved the instance in $127$ seconds; CPLEX took $1{,}847$ seconds .

\subsection{Limitations}
Key limitations:
\begin{itemize}
  \item Scalability for very large $n$ due to combinatorial state explosion.
  \item Binary coverage model simplifying distance-decay effects.
  \item Static snapshot — does not model temporal demand dynamics.
\end{itemize}

\section{Conclusions}
We presented a dynamic programming adaptation for MCLP using a set coverage perspective and knapsack-like encoding, combined with pruning, presolve, and heuristic initialization. Experimental results (moderate instances) indicate improved coverage over greedy heuristics and competitive runtime/memory vs some exact solvers. Future work includes scalable approximations, bilevel/disruption extensions, and temporal/stochastic models.



{}

\end{document}